\documentclass[
]{ceurart}

\usepackage[most]{tcolorbox}
\usepackage{xcolor}
\usepackage{needspace}
\usepackage{hyperref}

\usepackage{geometry} 
\usepackage{float}
\setcitestyle{authoryear,round}  
\definecolor{promptback}{rgb}{0.96, 0.96, 0.96}  
\definecolor{promptframe}{rgb}{0.80, 0.80, 0.80} 

\lstdefinestyle{promptstyle}{
  basicstyle=\itshape,    
  breakatwhitespace=false,
  breaklines=true,        
  captionpos=b,           
  extendedchars=true,     
  showspaces=false,       
  showstringspaces=false, 
  showtabs=false,         
  escapeinside={\%*}{*)}, 
  literate={- \*}{--}2
}

\usepackage{listings}
\usepackage[all]{nowidow}
\usepackage{tabularx}
\begin{document}

\copyrightyear{2026}
\copyrightclause{Copyright for this paper by its authors.
  Use permitted under Creative Commons License Attribution 4.0
  International (CC BY 4.0).}



\title{AI-Ready Research Workflows in Computational Social Science: Lessons on Building a Shared Language for Interdisciplinary Collaboration}




\author[1]{Joan Giner-Miguelez*}[%
email=joan.giner@bsc.es
]

\author[1]{Alexandra Málaga}[%
]

\author[1]{Felipe Gómez-Cortés}[%
]

\author[1]{Adrian Carrascosa}[%
]

\author[1]{Mariona Coll~Ardanuy}[%
]

\author[1,3]{Andrés F. Castro-Torres}[%
]

\author[2]{Raül Sirvent}[%
]

\author[2]{Rosa M. Badia}[%
]

\author[1]{Clara Guasch}[%
]

\author[1]{Mercè Crosas}[%
]

\address[1]{Computational Social Science and Humanities, Barcelona Supercomputing Center, BSC-CNS, Barcelona, Spain}

\address[2]{Workflows and Distributed Computing, Barcelona Supercomputing Center, BSC-CNS, Barcelona, Spain}

\address[3]{Digital and Computational Demography Laboratory, Max Planck Institute for Demographic Research, Rostock, Germany}

\cortext[1]{Corresponding author.}

\conference{Preprint article}
\begin{abstract}
Artificial intelligence (AI) is gaining traction in the social sciences and humanities (SSH). However, adoption remains limited by technical barriers to high-performance computing (HPC), validation processes that lag behind AI's rapid progress, and reproducibility standards that most SSH teams cannot meet. Research workflows—common in the life sciences—address these problems via encoding and abstracting technical complexity into repeatable routines; yet, accounts of how to build them in SSH remain scarce. We report on a two-year effort to build a workflow that enables a Science and Technology Studies unit to query, analyze, and enrich OpenAlex—a database of some 460 million scholarly records—on the MareNostrum supercomputer, using methods ranging from large-scale bibliometrics to LLM-based classification. We found the main challenge was translating domain-specific research questions into engineering requirements — bridging two distinct methodological languages, with implications that were both organizational and technical. Organizationally, it meant adopting and adapting Agile to the research rhythm and pace, and reframing collaboration from a service arrangement to a co-design process. Technically, model-driven engineering was as valuable for collaboration as it was for automation; co-building the model facilitated both the creation of a shared vocabulary and the abstraction of HPC complexity. Finally, we highlight limitations we found in validation, reproducibility, and FAIR metadata — beyond what any single project can sustain — calling for coordinated, cross-institutional investment in the tooling and standards needed for AI-ready SSH workflows sustainable at scale.
\end{abstract}


\maketitle


\section{Introduction}

Advances in artificial intelligence (AI) and the increased availability of data are opening research possibilities across the social sciences and humanities (SSH). For example, recent work spans language models that analyze social-network behavior at scale \citep{fraxanet2026decade} and map long-term cultural and semantic shifts in historical corpora \citep{kozlowski2019geometry}. In computer vision, recent work includes detecting archaeological sites from satellite imagery and historical maps \citep{berganzo2023curriculum} and hybrid vision-language models that transcribe manuscripts in historical and regional variants \citep{coll2025evaluating}. These applications are emerging in an ever-growing interdisciplinary landscape, where success relies on combining domain expertise with computational methods—whether through close collaborations or the rise of hybrid research profiles.

Despite these advances, integrating modern AI into SSH remains challenging, particularly given the intensive computational requirements of state-of-the-art models. For many SSH researchers, the main bottlenecks have shifted from conceptual alignment to complex engineering problems. First, there remains a steep barrier to entry for high-performance computing (HPC), whose tooling and technical culture were not designed with social scientists and humanists in mind. Second, as AI is adopted by researchers who may be unfamiliar with established machine learning and data science principles, methodological failures—such as flawed train-test splits and overinterpreted results—have become more common \citep{calder2022use}. Generative AI adds to this problem, since its outputs vary from run to run, making replication even harder \citep{bail2024can}.  Third, reproducibility of effective results remains a challenge. While basic repositories exist, achieving true computational reproducibility is incredibly time-intensive. Without systemic academic incentives, it is frequently neglected because researchers prioritize limited time, are unfamiliar with what reproducibility requires in daily practice, and lack seamless tooling that makes reproducible workflows the path of least resistance \citep{trisovic2022large}.

While fields like the life sciences have met scaling challenges with community-wide workflow initiatives \citep{da2024workflows}, SSH still lacks comparable frameworks for deploying large-scale AI. Even where high-quality, reproducible tools exist—such as those from the Computational Humanities Research (CHR) community \citep{mcdonough2025mapreader}—scaling AI methods like LLM-based classification over hundreds of millions of records on HPC infrastructures still relies on ad hoc, localized pipelines. Integrating AI at this scale introduces distinct bottlenecks: teams must coordinate massive parallel model inference, optimize high-throughput database querying, and design rigorous validation frameworks for non-deterministic model outputs. Consequently, SSH lacks practical "blueprints" \citep{ahnert2023collaborative} for co-designing workflows that make both HPC and heavy AI pipelines accessible to interdisciplinary teams. Following experience-report traditions in the life sciences \citep{schackart2024detailed}, we document our workflow designed to bridge this gap. We focus on translating domain research into engineering requirements, managing AI and HPC complexity for researchers, and establishing rigorous, reproducible validation frameworks for AI outputs.

We report on the design and construction of a shared research workflow for a science and technology studies (STS) unit over OpenAlex \citep{priem2022openalex}, based on roughly two years of work at the Computational Social Science and Humanities (CSSH) Lab of the Barcelona Supercomputing Center. The workflow combines organizational practices and technical automations that shield researchers from low-level supercomputing configurations, enabling them to query, analyze, and enrich the corpus at scale. In building it, we realized that our primary task was not merely delivering a single study, but building a shared vocabulary across the domain-engineering spectrum—especially for the hybrid profiles that increasingly bridge this divide. We structure these organizational (Section 3) and technical (Section 4) dynamics around three key takeaways: first, adopting and adapting Agile to academic rhythms and framing collaboration as co-design rather than a service arrangement is essential; second, that model-driven engineering \citep{brambilla2017model} is as valuable for building a shared vocabulary as it is for automation; and third, that the systemic challenges of validation, reproducibility, and FAIR sharing far surpass what any single project can sustain, requiring coordinated institutional and community investment in tooling and standards \mbox{(Section~\ref{sec:beyond})}.

\section{Context and running case}
\label{sec:con}

The lab is organized into three groups—Computational Social Science, Computational Humanities, and Methods and Engineering for SSH—with the Science and Technology Studies (STS) unit, whose work we draw on throughout this paper, belonging to Computational Social Science. While the Methods and Engineering group acts as a "mesh" connecting domain-focused groups to the HPC infrastructure, researchers and engineers operate across a domain-engineering spectrum, and a growing share of hybrid profiles work across both, applying AI and data science methods to their own domain research. The lab's mission is twofold: to support individual research projects and, in doing so, to develop and mature reusable infrastructure for computational SSH research. This dual mandate—deliver a study and leave behind something the next study can build on—shapes both practices we report: how teams organize their collaboration (Section \ref{sec:org}) and how they automate the work (Section \ref{sec:auto}).

The Science and Technology Studies unit has worked with the Methods and Engineering group over roughly two years to build and iterate on a shared infrastructure over OpenAlex \citep{priem2022openalex}—an open catalog of 460 million scholarly records, 60 million with full text. That infrastructure has supported several of the unit's studies, including analyses of how academic collaboration is changing with the rise of AI \citep{chacua2026}, of the dominance of "development" as a concept in the social sciences \citep{atorres2026d}, and of how the use of AI methods has spread across scientific fields and world regions since the 1960s \citep{torres2026ai}.

We take this last study as the running example throughout the paper. Answering its central question required classifying around 220 million works using a mix of LLM-based classification and dictionary tagging, amounting to some 0.9 billion model inferences in total. At that scale, engineering choices become research choices: a model-size reduction, validated against a stratified, human-evaluated sample before full execution, and a parallelized inference strategy—together, these cut the computation from an estimated 9,000 GPU-hours to 92 hours on 96 A100 GPUs on MareNostrum 5, roughly a hundredfold. Figure \ref{fig:workflow} shows its main findings: the prevalence of AI methods across the fields of science and their geographical distribution.

\begin{figure}[b]
  \centering
  \includegraphics[width=1\linewidth]{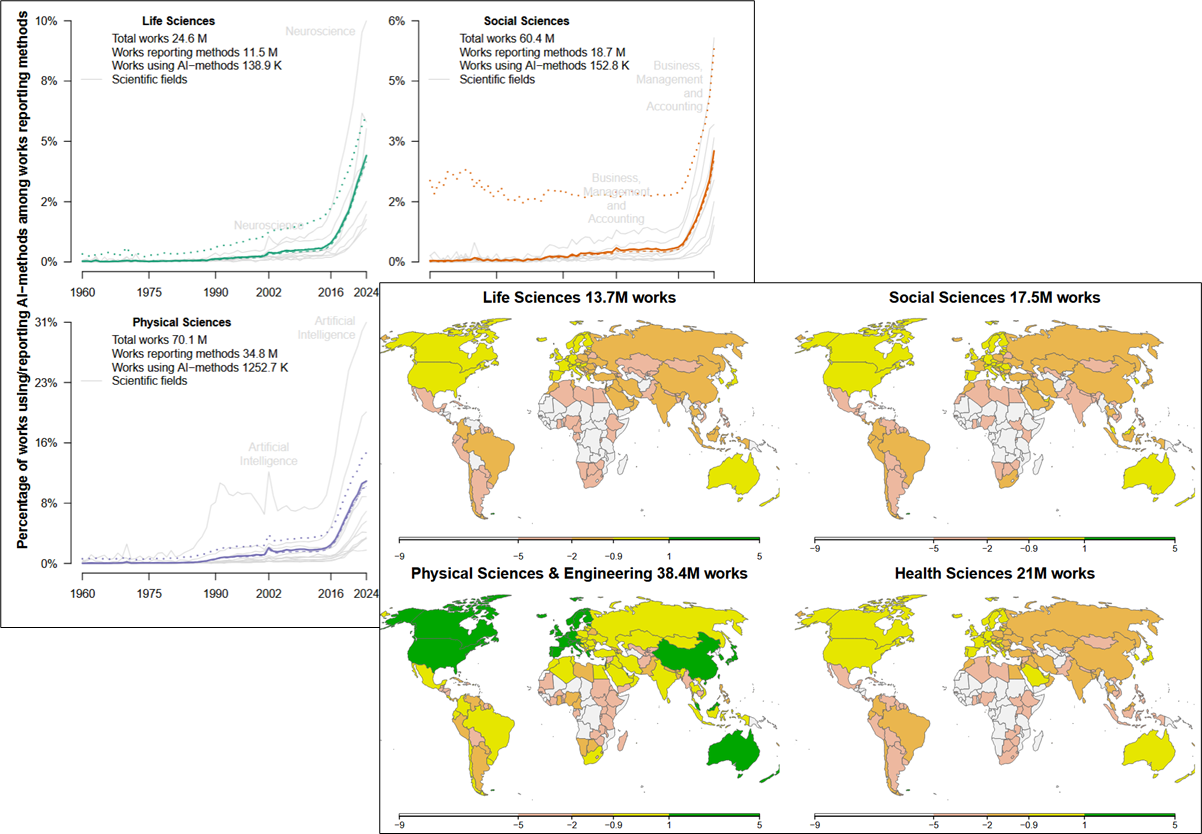}
  \caption{Results of the study about the prevalence of AI methods across scientific fields and world regions, produced by classifying ~220M OpenAlex records.}
  \label{fig:result}
\end{figure}

\section{Organising the collaboration}
\label{sec:org}

Assembling a team across the domain-engineering spectrum and getting its members to work as one was among the first organizational challenges we faced. In line with the Living with Machines project \citep{ahnert2023collaborative}, we found the central task to be building a shared "contact language" and the spaces for discussion in which to use it. This section presents our organizational takeaways from a series of retrospective meetings that brought together the core team, group directors, and principal investigators from other units where the workflow is now being adopted, while Section~\ref{sec:auto} shows how the shared language construction drove the technical pipeline design.


\textbf{Adopting and adapting Agile to research.} We adopted Agile — sprints, a shared JIRA backlog, sprint reviews — valuing it less for its delivery machinery than for the rhythm and shared vocabulary it imposed: a recurring occasion to surface misunderstandings early, before they compound. It cannot be imported at scale, and domain scientists bear the weight of the adaptation, as reshaping messy, non-linear inquiry into committed deliverables and user stories is a significant extra workload. Hybrid profiles, which move more fluidly across the domain-engineering spectrum, helped smooth this pressure without eliminating it. Pressed too rigidly, this project-management logic can overburden scientific work and delay or preclude discoveries by conflating means—tickets, epics, sprints, dashboards—with ends. We therefore kept Agile's structure but treated its commitments as revisable, following a process that is "iterative, self-reflexive, and designed to evolve" \citep{ahnert2023collaborative}—that is, the team regularly revisited not just what it delivered, but how it worked. Adapting Agile to a research environment demands this flexibility — and developing Agile variants better suited to research is, we think, a direction worth pursuing.

\textbf{Spaces for translating user stories.} The most concrete instrument was requirements translation: the shift from a domain-specific research question or need to the specific technical requirements it involves is where most interdisciplinary friction accumulates \citep{eisty2025ten}. Translating this into a first version of a ticket was itself a team effort, written in domain-specific needs terms (a.k.a user story: "as a historian, I need...") — rather than in engineering terms, from which specifications, deliverables, and acceptance criteria were then jointly derived. This translation was possible because the team had already invested in shared spaces and a common vocabulary, eased further by a growing share of hybrid profiles who move fluidly between domain and engineering framings. That investment falls as the team matures, and both organizational (the ticketing system) and technical tooling (the metamodel of Section ~\ref{sec:auto}) can help in sustaining it.

\textbf{From service to co-design approach.}  The relationship did not move in a single direction: it began close to a co-design approach, then shifted toward service alongside the adoption of Agile and JIRA tooling, before the team again pushed to restore co-design. How far engineers engage tracks task uncertainty: routine plumbing — such as scripting a standard database query — can be delegated, but when the method itself is an open question, engineering decisions become research decisions. For instance, in the LLM-based classification pipeline described in Section 2, validation techniques such as prompt-perturbation checks — proposed by team members with more engineering-focused profiles — tested the classification method's semantic consistency and strengthened the study's conclusions. This process lets us start a discussion at the lab level about the roles within the team — what, in fact, counts as a scientific contribution, and how such contributions should be credited.

\needspace{6\baselineskip} 
\section{Automating the workflow: a co-designed, model-driven vocabulary}
\label{sec:auto}

At OpenAlex scale, repeatable computation — pulling and cleaning data, deriving variables, running AI methods — hides considerable machinery: parallel execution, large-scale data management, AI model deployment. Our pipeline packages these automations with a twofold goal: to capture a use case's recurring elements in a shared conceptual space where the team can reason in common terms, and to let domain scientists run corpus-scale analyses without touching that machinery.

To build that space, we adopted model-driven engineering (MDE) \citep{brambilla2017model}, which makes an explicit model of a domain (a metamodel: its core concepts and their relations) the primary artifact for reasoning about the system, rather than raw code. We chose it less for its technical benefits than for communication: modeling is itself a conversation, and building the metamodel with the domain scientists — naming the workflow's pieces, agreeing how they fit — is a co-design process that produces a compact, shared vocabulary for reasoning about the use case. These are exactly the negotiations that, left implicit, produce the friction of Section \ref{sec:org}; making them explicit turns that shared vocabulary from an organizational aspiration into a concrete technical surface — the model-driven approach is as much an organizational instrument as a technical one.

\begin{figure}[b]
  \centering
  \includegraphics[width=0.9\linewidth]{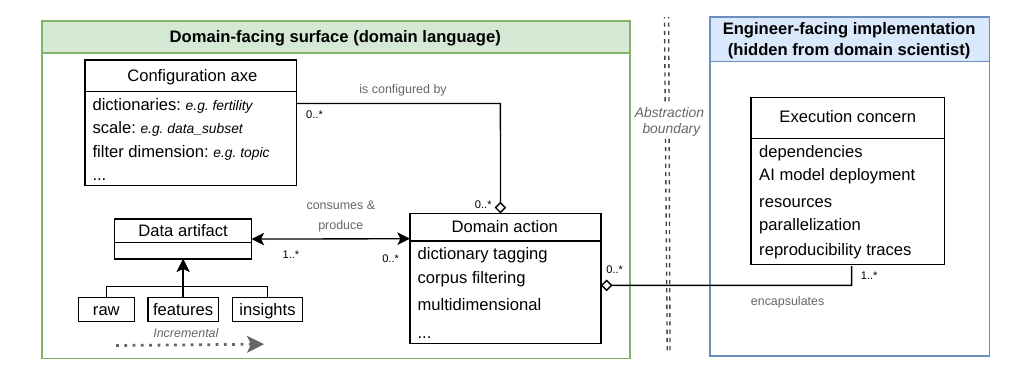}
  \caption{The metamodel, split by the abstraction boundary that divides the domain-facing surface (left) from the engineer-facing implementation (right). The four headers — incremental data artifacts, domain actions, configuration axes, and execution concerns — constitute the reusable vocabulary; the items within each box are their instantiations for the STS-over-OpenAlex case. A domain scientist works only on the left; everything on the right is encapsulated by the actions.}
  \label{fig:workflow}
\end{figure}

Figure~\ref{fig:workflow} shows the resulting four concepts. This vocabulary was not designed up front but distilled over two years of building the pipeline alongside the STS team and through retrospective meetings where we detected recurring operations and requests that the domain scientist repeatedly made, lifting each into a named, configurable action and packaging the execution concerns for each into reusable launchers. That method, more than the pipeline itself, is what we believe can help other projects build their own shared languages: the same approach is already being adapted for a Computational Humanities workflow transcribing ancient manuscripts \citep{coll2025evaluating}, suggesting it generalizes beyond the STS use case that motivates this paper.

\textbf{Incremental data artifact.} Data artifacts are layered: raw artifacts hold the source corpus; feature artifacts are enhanced outputs (e.g., country aggregations); insight artifacts are the subsets and aggregates that answer a research question. Reuse is the point — an expensive lower layer is computed once and serves many analyses above it — and each layer is versioned independently, so any analysis traces back to the exact versions that produced it, guarding against the inconsistent data snapshots that are a common source of methodological failure.

\textbf{Domain action.} Actions, the vocabulary's central piece, ingest and generate data artifacts, each exposed to the domain scientist as a single command — building topic features, aggregating author countries, or compacting the corpus — with the implementation hidden behind it. The representative case is dictionary tagging: a scientist writes a dictionary of words for a study — fertility, child labor, our running case's AI methods — and tags every abstract in the corpus against it with one command, returning a new column plus a reproducibility trace. The same validation protocol can be attached once and reapplied on every run, guarding against the task-framing and evaluation errors flagged earlier. Swapping the dictionary serves a different study: what the scientist controls is domain language, not code.

\textbf{Configuration axe.} These are the dimensions domain scientists configure within an action: the scale of the operation (a small subset for iteration, or the entire corpus), filters on any feature in the data layers, or — as in the dictionary tagging example — the specific dictionary defining a study's topic, e.g., fertility.

\textbf{Execution concern.} Everything below the boundary — software environments, resource requests, parallelization, model deployment — is absorbed by a small set of reusable launchers. An action binds a source module to a launcher, which handles job submission, runs COMPSs \citep{badia2026programming} to distribute work across nodes, and emits per-task logs consolidated into a reproducibility record — letting the STS team treat a corpus-scale job as a single command, not an HPC project.

The natural next step, we think, is to lean further into MDE practices and move toward a domain-specific language with code-generation capabilities, letting per-project models evolve faster and more safely than conventions allow and opening the door to low-code and visual interfaces \citep{alfonso2024building}, making research workflows even more accessible.

\section{Beyond results: trustworthy and reusable outputs}
\label{sec:beyond}

During the project, we achieved only partial success in delivering trustworthy, reproducible, and reusable results: validation was under-resourced, reproducibility was only partially automated, and metadata was largely produced by hand. Each of these limitations, we found, was largely beyond what a single project could close on its own; in this section, we recast each as a takeaway — for institutional, cross-institutional, and community-wide efforts — that future projects will need to succeed.

\textbf{Institutional infrastructure to power human validations.}  Validation has long been central to NLP research — not a final check, but part of the scientific work itself. As AI spreads across SSH, and especially with general-purpose LLMs, this need only sharpens: we increasingly apply LLMs to tasks they were never evaluated on — classifying scientific abstracts, as in our own running case, or studying social behavior — with no benchmark to tell us how they'll perform or what biases they bring. For now, that gap can only be closed with costly human labor. We under-anticipated this cost: the human evaluation of a stratified sample was designed only once the outputs were ready, and coordinating enough qualified evaluators delayed the results. Validation must therefore be budgeted and scheduled from the start. This is precisely why validation surpasses what any single project can sustain: an institution needs standing capacity — crowdworkers, expert partners, or internal staff — that it can pull on as smoothly as it already does compute infrastructure, rather than each project building that capacity from scratch under deadline pressure.

\textbf{Composable reproducibility artifacts to match research processes.} Packaging a result to be re-run is, in principle, a natural output of the automation of Section~\ref{sec:auto}: COMPSs can already emit a reproducibility capsule (RO-Crate \citep{soiland2022packaging}) for a single action~\citep{sirvent2022automatic}. In practice, progress was partial. A domain task usually spans several actions, and composing their action-level records into hierarchical RO-Crates that describe the whole task — harder still when applications are nondeterministic~\citep{sirvent2025reproducibility} — relies on manual effort, as automation does not yet exist. Closing that gap is not any single project's to solve — it is an open research direction for the wider reproducibility and standards community.

\textbf{Shared tooling for findable, reusable metadata.} The last step is describing the dataset for discovery and reuse — a Croissant~\citep{akhtar2024croissant} description the ML community consumes directly, and DDI-CDI \footnote{DDI-CDI specification homepage: \url{https://ddialliance.org/ddi-cdi}} metadata tying it into the wider social-science infrastructure. Here, progress was thinnest: community tooling was still immature—our Dataverse \citep{king2007introduction} instance was mid-deployment, DDI-CDI integration required custom work we could not generalize, and Croissant support was limited—so we produced it largely by hand. Maturing this ecosystem (better Dataverse support, generators such as Croissant Baker~\citep{attrach2026croissant} is a shared opportunity: the tooling is the same for every SSH team, so labs are better off building it together than reinventing it individually.

\section{Conclusions}

Two years of building an AI-ready research workflow for a Science and Technology Studies (STS) unit taught us the organizational and technical nuances of building a shared vocabulary and collaborative spaces for interdisciplinary AI research. Organizationally, three things helped bridge that gap: adapting and adopting Agile to the rhythm of research; pushing our own collaboration — which moved between service and co-design more than once — back toward co-design; and the emergent role of hybrid profiles, who eased friction that would otherwise have fallen entirely on domain scientists. Technically, model-driven engineering (MDE) proved just as effective at fostering collaboration as it was at automating processes, and we plan to expand our adoption of MDE in future work. However, significant systemic barriers remain in building AI-ready research workflows that consistently meet best practices. First, institutions should develop dedicated infrastructure to poll experts and manage validation teams. Second, current reproducibility tooling and standards should be adapted to support highly composable research processes. Finally, achieving true sustainability requires a coordinated, cross-institutional effort to cultivate a mature tool ecosystem that can automatically generate machine-actionable FAIR metadata.

\section*{Data Accesiblity Statement}

The code for the presented pipeline \citep{giner_miguelez_2026_21512187} and the data from the STS unit study \citep{4HDU4Y_2026}, are openly available at the BSC Dataverse instance.

\begin{acknowledgments}
Joan Giner-Miguelez, Alexandra Málaga, Adrián Carrascosa, and Mariona Coll Ardanuy acknowledge their AI4S fellowship within the “Generación D” initiative by Red.es, Ministerio para la Transformación Digital y de la Función Pública, for talent attraction (C005/24-ED CV1), funded by NextGenerationEU through PRTR. Andres F. Castro acknowledges funding from Grant RYC2023-042730-I,by MICIU/AEI/10.13039/501100011033andESF+. This paper has been partially supported by the projects CEX2021-001148-S, and PID2023-147979NB-C21 from the MCIN/AEI and MICIU/AEI /10.13039/501100011033 and by FEDER, UE, by the Departament de Recerca i Universitats de la Generalitat de Catalunya, research group MPiEDist (2021 SGR 00412).
\end{acknowledgments}



\end{document}